# First-principles predictions of HfO$_2$-based ferroelectric superlattices

Binayak Mukherjee[1], Natalya S. Fedorova[1] and Jorge Íñiguez-González[1,2]

[1]Department of Materials Research and Technology, Luxembourg Institute of Science and Technology, L-4362, Esch-sur-Alzette, Luxembourg

[2]Department of Physics and Materials Science, University of Luxembourg, L-4422, Belvaux, Luxembourg

The metastable nature of the ferroelectric phase of HfO$_2$ is a significant impediment to its industrial application as a functional ferroelectric material. In fact, no polar phases exist in the bulk phase diagram of HfO$_2$, which shows a dominant non-polar monoclinic ground state. As a consequence, ferroelectric orthorhombic HfO$_2$ needs to be kinetically stabilized. Here, we propose an alternative approach, demonstrating the feasibility of thermodynamically stabilizing polar HfO$_2$ in superlattices with other simple oxides. Using the composition and stacking direction of the superlattice as design parameters, we obtain heterostructures that can be fully polar, fully antipolar or mixed, with improved thermodynamic stability compared to the orthorhombic polar HfO$_2$ in bulk form. Our results suggest that combining HfO$_2$ with an oxide that does not have a monoclinic ground state generally drives the superlattice away from this non-polar phase, favoring the stability of the ferroelectric structures that minimize the elastic and electrostatic penalties. As such, these diverse and tunable superlattices hold promise for various applications in thin-film ferroelectric devices.

## 1. Introduction

Originally studied as a high-permittivity dielectric with commercial applications in the mass production of complementary metal-oxide-semiconductors (CMOS), it is only relatively recently that HfO$_2$ has been established as a ferroelectric (FE), with the first reports appearing in 2011[1,2]. Subsequently, FE HfO$_2$ has become the subject of intense research interest – not just for its CMOS-compatibility and associated technological promise, but also for the atypical origin and nature of ferroelectricity in this fluorite-structured simple oxide. Multiple theoretical models have been proposed[3–6], suggesting both an improper[3] and proper[4,5] nature of ferroelectricity, with recent experimental results favoring the latter[7]. The piezoresponse of HfO$_2$ is just as peculiar, displaying a potentially tunable[8] longitudinal piezoelectric effect whose sign is non-trivial[9–12].

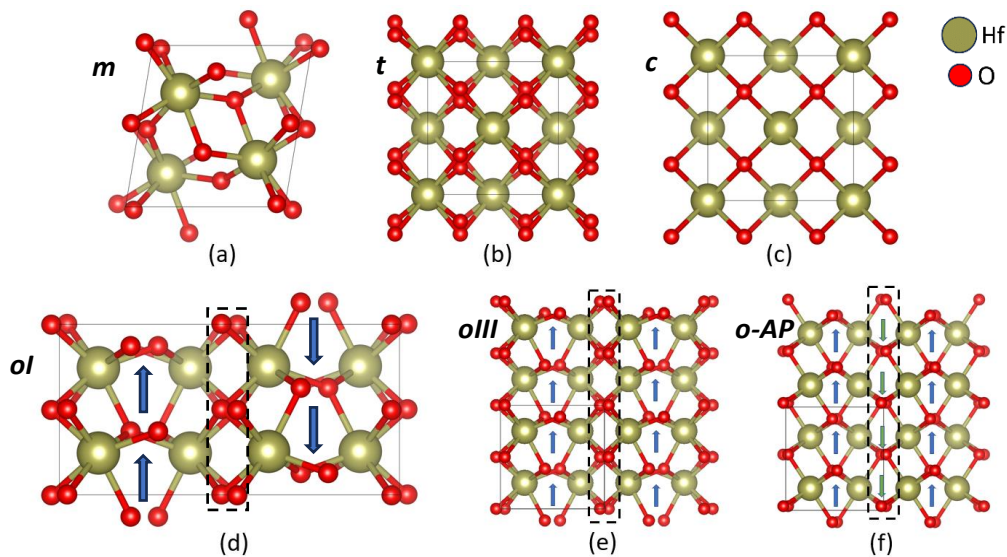

Figure 1. Some relevant polymorphs HfO$_2$, including the (a) monoclinic, (b) tetragonal, (c) cubic, (d) orthorhombic-I, (e) orthorhombic-III, and (f) orthorhombic-AP phases. The phase label is given in the top left corner of each structure. The arrows indicate the direction of the local dipoles in each half unit cell, taking the cubic phase (panel d) as reference; the dashed boxes identify the so-called "spacer layers".

The ground state for bulk HfO$_2$ at room temperature and pressure is the monoclinic "m" phase (space group *P2$_1$/c*, Fig. 1a). At 1973 K, it transitions into the tetragonal "t" phase (*P4$_2$/nmc*, Fig. 1b), and then at 2773 K into the cubic "c" phase (*Fm-3m*, Fig. 1c). With increasing pressure instead, the m phase transitions into an antipolar orthorhombic "oI" phase (*Pbca*, Fig. 1d), and then into a different orthorhombic "oII" phase (*Pnma*). The polar phases, which do not appear in the bulk phase diagram, include the rhombohedral (*R3m*) and orthorhombic "oIV" (*Pmn2$_1$*) states, as well as the most common ferroelectric orthorhombic "oIII" phase (*Pca2$_1$*, Fig. 1e)[1]. The polarization in this oIII phase is quite distinct from that in ferroelectric perovskites and can be visualized as an off-centering of half the oxygens (which we will call "polar" or "active" oxygens in the following, marked by blue arrows in Fig. 1e) in the unit cell, with respect to the high-symmetry positions they occupy in the cubic structure.[13] By contrast, the other half of the oxygens ("spacer" or "buffer" oxygens in the following, dashed box in Fig. 1e) remain close to their high-symmetry positions and do not contribute significantly to the polarization. The oIII phase presents an oxygen coordination of 7 for each cation (3 polar and 4 buffer), as described by the so-called 7C theory of ferroelectricity in HfO$_2$, which claims that this unusual coordination number serves as a fingerprint of ferroelectricity in simple oxides and halides[14]. A closely related polymorph of interest is the higher energy antipolar "o-AP" phase (*Pbcn*, Fig. 1f), where we may say that the buffer oxygens undergo an off-centering equal and opposite to that of the polar oxygens (dashed box and green arrows in Fig. 1f). This phase is one among several proposed paraelectric reference structures to discuss ferroelectricity in the oIII phase[4,5], but – as far as we know – has never been observed experimentally in bulk HfO$_2$.

The stabilization of the metastable FE oIII phase has been attributed to a multitude of factors, including oxygen vacancies, dopant species and concentration, surface energy minimization, quenching kinetics, and mechanical effects[15]. In fact, both conventional nucleation theory[16] as well as atomistic nudged elastic band (NEB) calculations using density functional theory (DFT)[17,18] suggest that the oIII phase is stabilized kinetically over the m phase, which is the thermodynamic ground state. As such, replacing the monoclinic structure with a polar ground state remains an open problem.

A promising way forward is through the design of nanostructures that may allow us to achieve that goal. One such candidate nanostructure is a superlattice (SL), i.e., a periodic lattice consisting of nanometric layers of two different materials. DFT calculations have shown that Si dopants in oIII HfO$_2$ adopt stable, layered configurations akin to SLs, and lead to a ferroelectric oIII ground state[19]. Recent reports on ZrO$_2$/HfO$_2$ (Zr/Hf) SLs demonstrate enhanced polarization, improved reliability at high temperatures, a tunable coercive field[20] and high stability of the ferroelectric state[21]. Such superlattices have recently been used as gate dielectrics in transistors[22] and to obtain wake-up free ferroelectric capacitors[23]. Despite the heightened interest, the mechanism behind the formation of the oIII phase in the superlattices is not fully understood, though some propose that it is connected to the in-plane tensile strain at the interface[24]. A DFT study has also suggested that the enhanced endurance of these FE SL's can be explained by a suppression of oxygen vacancies[25].

The promising results obtained for Zr/Hf SLs naturally justify an interest in similar heterostructures with other simple oxides. Yet, such studies are conspicuous by their absence, a situation which the present work seeks to partially remedy. Here we report our DFT results on superlattices of HfO$_2$ with a series of simple oxides, predicting that we can thus stabilize ground state phases with polar, antipolar, and mixed polar/antipolar or polar/nonpolar characters. Our results lend themselves to a simple interpretation in terms of elastic and electrostatic considerations, and allow us to identify the most promising directions for the growth of ferroelectric $HfO_2$-based superlattices

## 2. Computational approach

### 2.1. Pure compounds

To explore the behavior of the SL's, we first consider the bulk structures of the pure compounds. Apart from $HfO_2$, we choose all the other group IV (A and B) oxides, as well as the fluorite structured lanthanide $CeO_2$. These are all simple compounds with the chemical formula $XO_2$, where X is a 4+ cation (X = Si, Ge, Ti, Sn, Zr, Pb and Ce, in increasing order of cation radius, Table 1). They are mostly ionic in nature with large band gaps. Starting from geometries isostructural to the oIII, oI, m and t phases of $HfO_2$, each pure compound is fully relaxed to determine the relative stability of such polymorphs, leading to an interesting structural classification.

Table1. The chosen oxides, their cation radii[26], selected polymorphs considered in this work, and their corresponding energies relative to their respective ground state.

| Oxide | Cation radius (Å) | Structure (*space group*) | Energy (meV/cation) |
|---|---|---|---|
| $SiO_2$ | 0.4 | *I-42d* | 0 |
|  |  | *Pbcn* (o-AP) | 346 |
|  |  | *P4$_2$/nmc* (t) | 702 |
|  |  | *Pbca* (oI') | 1335 |
| $GeO_2$ | 0.53 | *P4$_2$/mnm* | 0 |
|  |  | *Pbcn* (o-AP) | 98 |
|  |  | *P4$_2$/nmc* (t) | 618 |
|  |  | *Pbca* (oI') | 742 |
| $TiO_2$ | 0.74 | *Pbcn* (o-AP) | 0 |
|  |  | *P2$_1$/c* (m) | 28 |
|  |  | *Pbca* (oI) | 158 |
|  |  | *Pca2$_1$* (oIII) | 172 |
|  |  | *P4$_2$/nmc* (t) | 278 |
| $SnO_2$ | 0.81 | *P4$_2$/mnm* | 0 |
|  |  | *Pbcn* (o-AP) | 49 |
|  |  | *Pbca* (oI) | 329 |
|  |  | *P4$_2$/nmc* (t) | 433 |
| $HfO_2$ | 0.83 | *P2$_1$/c* (m) | 0 |
|  |  | *Pbca* (oI) | 46 |
|  |  | *Pca2$_1$* (oIII) | **64** |
|  |  | *Pbcn* (o-AP) | 127 |
|  |  | *P4$_2$/nmc* (t) | 139 |
| $ZrO_2$ | 0.84 | *P2$_1$/c* (m) | 0 |
|  |  | *Pbca* (oI) | 41 |
|  |  | *Pca2$_1$* (oIII) | 52 |
|  |  | *P4$_2$/nmc* (t) | 79 |
| $PbO_2$ | 0.94 | *Pbcn* (o-AP) | 0 |
|  |  | *Pbca* (oI) | 56 |
|  |  | *P4$_2$/nmc* (t) | 102 |
| $CeO_2$ | 0.97 | *Fm-3m* | 0 |

It is noteworthy that *only* HfO$_2$ and ZrO$_2$ are found to have ground states in the centrosymmetric m phase, whereas for SiO$_2$, GeO$_2$, SnO$_2$ and PbO$_2$ neither the m phase nor the oIII phase are stable. Starting from either of these phases and minimizing the Hellmann-Feynman forces and total energy causes the systems to relax into the centrosymmetric antipolar o-AP structure (Fig 1f). This structure is also the ground state for PbO$_2$[27] and TiO$_2$[28], with the latter additionally having local minima corresponding to the m and oIII phases. GeO$_2$ and SnO$_2$ have ground states in the structurally similar tetragonal rutile *P4$_2$/mnm* phase[29,30], while the ground state for SiO$_2$ has been proposed to be the tetragonal *I-42d* phase[31]. The oI phase, which is the preferred antipolar configuration for HfO$_2$ and ZrO$_2$, also exists as a high energy state in the other oxides, though in a distorted form (oI') in SiO$_2$ and GeO$_2$ (see supplementary note S1), alongside higher energy tetragonal phases. Finally, we have the peculiar case of CeO$_2$, which strongly favors a ground state in the fluorite c phase and relaxes to that structure regardless of the starting geometry, again suggesting the absence of any oIII or m local minima.

## 2.2. Superlattices

For the purposes of this study, we construct SLs with infinitely repeating, alternating layers of HfO$_2$ and XO$_2$. Unless otherwise indicated, in our simulations the starting geometries for the SLs are the same for both layers, and four separate polymorphs are considered: the polar oIII phase, and the centrosymmetric oI, m, and t phases. The initial structures are allowed to relax fully, without imposing any epitaxial conditions. (This would correspond to free-standing or fully-relaxed films in experiments.) Two layer thicknesses are considered, with either 2 or 4 cation sublayers in each layer, denoted as 2/2 and 4/4 SLs respectively. The SLs are stacked along the pseudo-cubic [100] (A-axis), [010] (B-axis), and [001] (C-axis) directions, and are respectively called A-, B- and C-SLs. We define our frame of reference such that the polarization in oIII HfO$_2$ is along the C-axis (Fig. 2a). The A and B directions, both perpendicular to the polarization, can be distinguished by the orientation of the spacer layer, which lies parallel to the B-axis and perpendicular to the A-direction (Fig 2b,c). Note that this frame of reference is uniquely defined for the oIII and m phases, while for the t phase the A and C directions (perpendicular to the 4-fold axis) are equivalent. Furthermore, the oIII and m SLs stacked along B can have the interface passing either through the polar layer (B$_{pol}$) or the spacer layer (B$_{spc}$), allowing for two inequivalent geometries (Figs. 2c and d respectively).

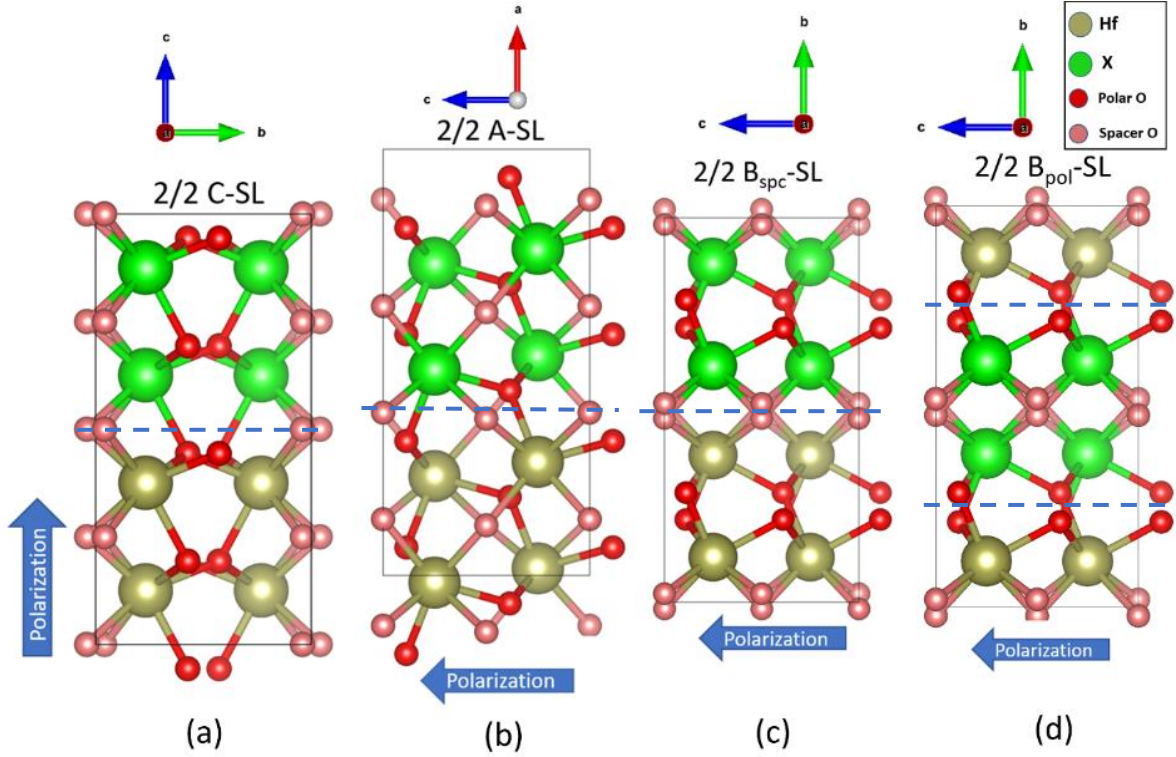

Figure 2. Sketches of the considered 2/2 X/Hf oIII SL's superlattices: (a) stacking in the C-direction (parallel to polarization), (b) stacking in the A-direction (perpendicular to polarization, mixed interface), (c) stacking in the B-direction (perpendicular to polarization, spacer interface), and (d) stacking in the B-direction (perpendicular to polarization, polar interface). Dashed line demarcates the interface.

After each SL is completely relaxed, its formation energy is computed. For an X/Hf superlattice with equally thick XO$_2$ and HfO$_2$ layers, in some phase $p$ (with $p$ = oIII, oI, m, t), the formation energy per cation is defined as,

$$\Delta E^{SL}_{p-X,Hf} = E^{SL}_{p-X,Hf} - \frac{1}{2}\left(E^{Bulk}_{gs-XO_2} + E^{Bulk}_{gs-HfO_2}\right) \quad (1)$$

Here, $E^{SL}_{p-X,Hf}$ is the cohesive energy per cation of an X/Hf SL in phase $p$. Additionally, $E^{Bulk}_{gs-XO_2}$ and $E^{Bulk}_{gs-HfO_2}$ are the cohesive energies per cation of the fully relaxed pure compounds in their respective ground state. The second term on the right is then just the average of the bulk ground state cohesive energies of the pure compounds, and the difference with $E^{SL}_{p-X,Hf}$ gives the energy cost of producing the superlattice with respect to the separate bulks.

In order to discuss the relative stability of a particular superlattice with respect to the lowest-energy monoclinic (m) or distorted monoclinic (m') SL of the same composition, we introduce a quantity denoted "energy penalty", which is simply the difference in formation energy of the two SLs. More specifically, for an X/Hf SL in some phase $p$, the energy penalty with respect to its corresponding lowest-energy m (or m') SL is defined as,

$$\Delta E^{pen,SL}_{p-X,Hf} = \Delta E^{SL}_{p-X,Hf} - \Delta E^{SL}_{m-X,Hf} \quad (2)$$

Any SLs whose energy penalty is less than the energy of bulk polar oIII HfO$_2$ ($\Delta E^{pen,SL}_{p-X,Hf} <$ 64 meV/cation, see Table 1) are more likely to support the ferroelectric state than bulk HfO$_2$ is. These SLs are hereafter referred to as 'competitive'. Furthermore, the SL's that satisfy the condition $\Delta E^{pen,SL}_{p-X,Hf} <$

0 correspond to a case where the phase *p* becomes the ground state over the m or m' SLs. These SL's are hereafter referred to as 'favorable'.

The formation energy of a particular phase of a particular SL is naturally the consequence of multiple, competing structural features and physical mechanisms. Two of these can be easily identified – the contributions from the bulk structure of each individual layer, and from the elastic constraints imposed at the interface of the two layers. Indeed, we can express the formation energy of an X/Hf SL in some phase *p* as the sum of a bulk contribution, an elastic contribution, and the remaining non-elastic contributions coming from the interfacial discontinuity:

$$\Delta E_{p-X,Hf}^{SL} = \Delta E_{p-X,Hf}^{Bulk\ avg} + \Delta E_{p-X,Hf}^{Elastic} + \Delta E_{p-X,Hf}^{Non-elastic} \quad (3)$$

The first term is an average of the energies of the pure compounds ($XO_2$ and $HfO_2$) in the phase *p*, defined with respect to their corresponding ground state energies:

$$\Delta E_{p-X,Hf}^{Bulk\ avg} = \frac{1}{2}\left[\left(E_{p-XO_2}^{Bulk} - E_{gs-XO_2}^{Bulk}\right) + \left(E_{p-HfO_2}^{Bulk} - E_{gs-HfO_2}^{Bulk}\right)\right] \quad (4)$$

where $E_{p-XO_2}^{Bulk}$ and $E_{p-HfO_2}^{Bulk}$ are the cohesive energies of pure $XO_2$ and $HfO_2$ in phase *p*. The elastic contribution is instead given by

$$\Delta E_{p-X,Hf}^{Elastic} = \frac{1}{2}\left[\left(E_{p-XO_2}^{Const\ bulk} - E_{p-XO_2}^{Bulk}\right) + \left(E_{p-HfO_2}^{Const\ bulk} - E_{p-HfO_2}^{Bulk}\right)\right] \quad (5)$$

where $E_{p-XO_2}^{Const\ bulk}$ and $E_{p-HfO_2}^{Const\ bulk}$ are the cohesive energies of the pure compounds as obtained from a constrained relaxation, wherein we keep their in-plane lattice parameters strained to match those of the corresponding SL's. This term quantifies how much of the SL formation energy comes purely from the elastic straining of the pure compounds when they are put together in a heterostructure.

Finally, substituting (1), (3), and (4) into (2), we obtain the non-elastic component:

$$\Delta E_{p-X,Hf}^{Non-elastic} = E_{p-X,Hf}^{SL} - \frac{1}{2}\left(E_{p-XO_2}^{Const\ bulk} + E_{p-HfO_2}^{Const\ bulk}\right) \quad (6)$$

which is essentially a catch-all term that accounts for all effects due to the interface which are *not* explicitly elastic. These contributions, which are difficult to isolate, include chemical effects at the interface such as changes in bonding, as well as electrostatic effects due to depolarizing fields.

## 3. Results and discussion

### 3.1. Energetics

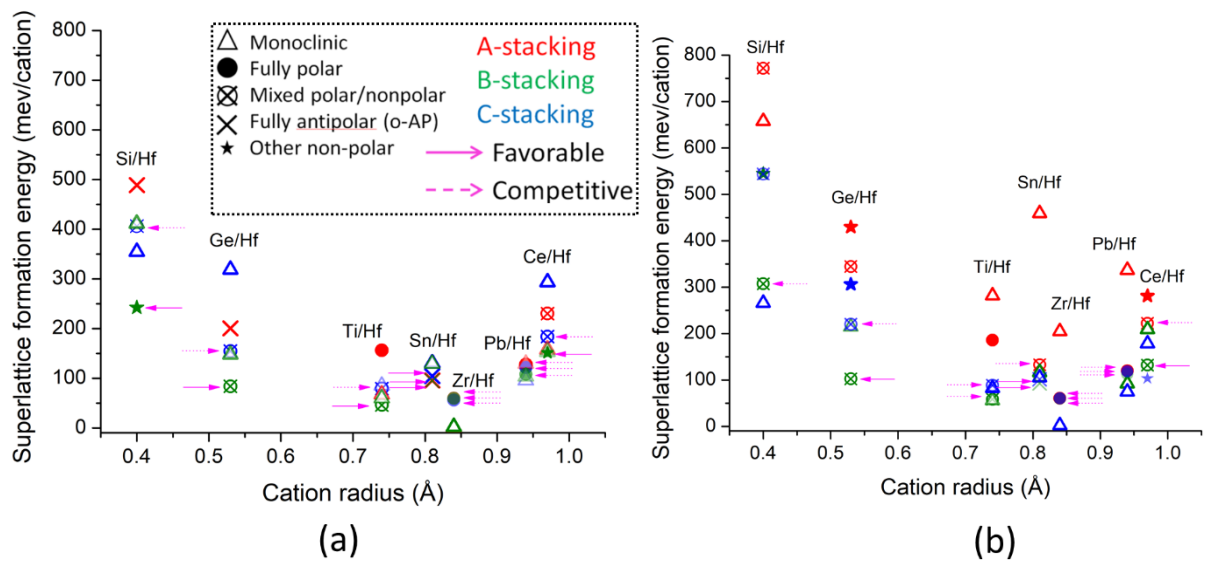

Figure 3. Formation energies of relevant SL's with monoclinic, polar, antipolar and mixed configurations for (a) 2/2 and (b) 4/4 systems, plotted as a function of the radius of the non-Hf cation The dashed arrows identify phases in the competitive range, while the solid arrows indicate the favorable SL's.

The formation energies for some 2/2 and 4/4 SLs are shown in Figs. 3(a) and 3(b), respectively, plotted as a function of the ionic radius of the X cation. Included are all the SLs relaxed starting from oIII and leading to various polar, nonpolar and mixed configurations, as well as the m SLs that are their competitors. Large variations in formation energies are observed for different stacking directions and compositions. Depending on these design parameters, the SLs starting from oIII end up in either fully polar (Fig 3, '●'), fully antipolar (Fig 3, '✗'), or mixed polar/nonpolar (Fig 3, '⊗') geometries with polar $HfO_2$ and centrosymmetric $XO_2$ layers. Similarly, the initially monoclinic SLs generally relax to structures either close to the usual m phase or to a distorted m'-phase (both denoted by '△' in Fig 3), and in a few cases relaxes to entirely different phases (Fig 3, '★'). Our main focus will naturally be on the low-laying SLs obtained from the oIII starting configuration, with a twofold interest: (a) the differences in their formation energies vis-a-vis the m SLs, and (b) the net polarization of the relaxed SL's. As it turns out, we can identify 23 orthorhombic SL's that are competitive with m SL's, i.e. an improvement over bulk oIII $HfO_2$ (Fig 3, dashed arrows), as well as 4 polar and 7 antipolar configurations which are outright favorable (Fig 3, bold arrows). These details are summarized in Table 2, and further discussed in the subsequent sections. For the formation energies of the various SLs grouped by composition, refer to Supplementary Tables S1-S7.

Table2. Structures, energies, polarizations of (initially) oIII SL's after relaxation. Formation energies are given in meV/cation, polarizations in C/m$^2$. Structures are classified as cubic (C), polar (P), antipolar (A) and monoclinic (M) for the X/Hf layers. Yellow shading indicates competitive superlattices, light green indicates favorable non-polar structures, dark green indicates favorable polar structures, and red indicates high energy phases. The structures with the asterisk are pictorially shown in the next section.

| Composition | Stackings: Thickness: | A-SLs 2/2 | 4/4 | B-SLs 2/2 | 4/4 | C-SLs 2/2 | 4/4 |
|---|---|---|---|---|---|---|---|
| Zr/Hf | En. Penalty | 58 | 56 | 59 | 56* | 56 | 58* |
|  | Polarization | 0.56 | 0.56 | 0.56 | 0.56 | 0.55 | 0.55 |
|  | Structure | P/P | P/P | P/P | P/P | P/P | P/P |
| Pb/Hf | En. Penalty | 32 | 45 | 12 | 38 | 26 | 44 |
|  | Polarization | 0.47 | 0.47 | 0.53 | 0.54 | 0.52 | 0.51 |
|  | Structure | P/P | P/P | P/P | P/P | P/P | P/P |
| Sn/Hf | En. Penalty | -33 | 27 | -33 | -13* | -24 | -4* |
|  | Polarization | 0 | 0.2 | 0 | 0 | 0 | 0 |
|  | Structure | A/A | A/P | A/A | A/A | A/A | A/A |
| Si/Hf | En. Penalty | 134 | 506 | -28 | 41 | 52 | 278 |
|  | Polarization | 0 | 0.37 | -0.1 | 0.21 | 0.26 | 0.28 |
|  | Structure | A/A | A/P | A/A | A/P | A/P | A/P |
| Ge/Hf | En. Penalty | 56 | 130 | -63 | -112* | 7 | 6* |
|  | Polarization | 0 | 0.23 | 0.1 | 0.2 | 0.22 | 0.24 |
|  | Structure | A/A | A/P | A/P | A/P | A/P | A/P |
| Ti/Hf | En. Penalty | 97 | 129 | -13* | 45 | 20 | 32 |
|  | Polarization | 0.69 | 0.68 | 0.16 | 0.21 | 0.36 | 0.38 |
|  | Structure | P/P | P/P | A/P | A/P | A/P | A/P |
| Ce/Hf | En. Penalty | 77 | 44 | -2 | -46* | 31* | -75* |
|  | Polarization | 0.2 | 0.23 | 0 | 0.18 | 0.23 | 0 |
|  | Structure | C/P | C/P | C/A | C/P | C/P | C/M |

## 3.2. Most promising polar superlattices

Of the various configurations outlined in Table 1, we first discuss the compositions where a polar solution is energetically favorable, as predicted for the mixed Ge/Hf, Ti/Hf and Ce/Hf superlattices.

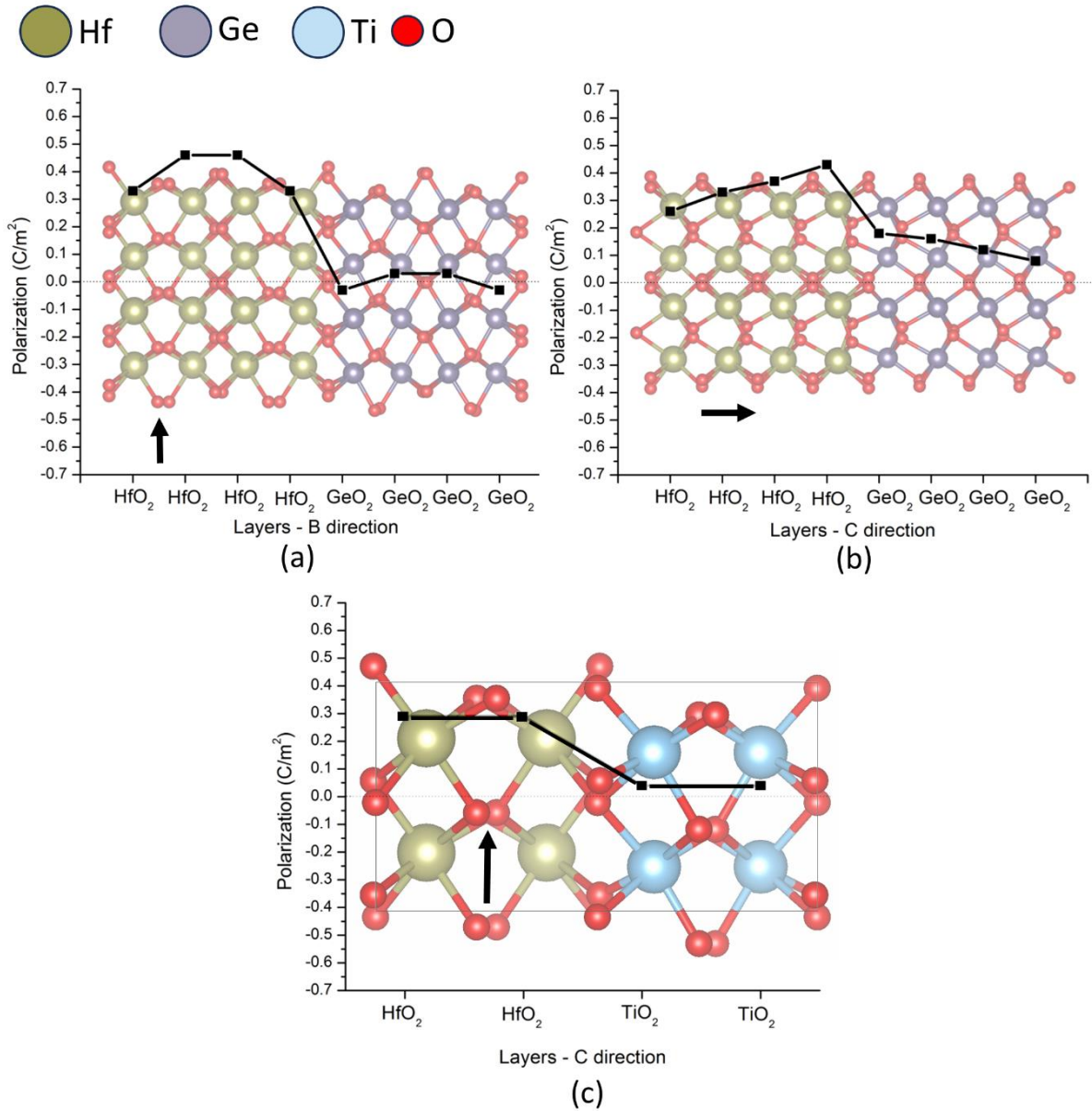

Figure 4. Structure and sublayer polarization of 4/4 Ge/Hf mixed SLs with (a) polarization in plane (B-SL) and (b) polarization out of plane (C-SL); (c) 2/2 Ti/Hf mixed SL with polarization in plane (B-SL); all structures were obtained by relaxing a fully oIII initial configuration.

The lowest energy configuration where a polar structure is stabilized is the B-oriented Ge/Hf system with 4/4 layer thicknesses (Fig 4a). In this SL, the $HfO_2$ layer remains close to the bulk oIII phase with the polarization in the plane of the interface, while the $GeO_2$ layer tends towards its o-AP low energy bulk polymorph. The sublayer polarization peaks in the middle of the $HfO_2$ layer, and then drops sharply across the interface to approximately zero in the $GeO_2$ layer. The 2/2 Ge/Hf B-SLs are also similarly polar and favorable.

Another interesting structural feature is observed for the energetically very competitive 4/4 mixed oAP/oIII C-SL, which supports out-of-plane polarization across the polar/antipolar interface (Fig 4b). In this case, the GeO$_2$ layer gets significantly polarized, so as to minimize the polarization discontinuity at the Ge/Hf interface and thus reduce the depolarizing fields. The polarization increases towards one interface and decreases at the other, suggesting the presence of bound charges at the interface. This is confirmed by the electronic density of states, which shows a decreasing band gap in the Ge/Hf C-SLs with increasing layer thickness (Supplementary Fig S1); however, a metallic interfacial state is not present in the relatively short-period SLs investigated here.

Crucially, these low-energy o-AP/oIII SLs also have lower energy penalties than the corresponding fully antipolar oI'/oI SLs. Additionally, since GeO$_2$ has a rutile ground state, fully rutile structured SLs as well as mixed rutile/m and rutile/oIII SLs were also evaluated, and found to have formation energies higher than their o-AP/oIII counterparts [Supplementary Table S3].

A further favorable configuration is obtained for the mixed Ti/Hf 2/2 o-AP/oIII B-SL (Fig 4c) with a structure and polarization very similar to those of the corresponding Ge/Hf B-SLs. Additionally, both 2/2 and 4/4 mixed C-SLs and the 4/4 mixed B-SLs are energetically quite competitive against the m SLs and structurally similar to their Ge/Hf counterparts.

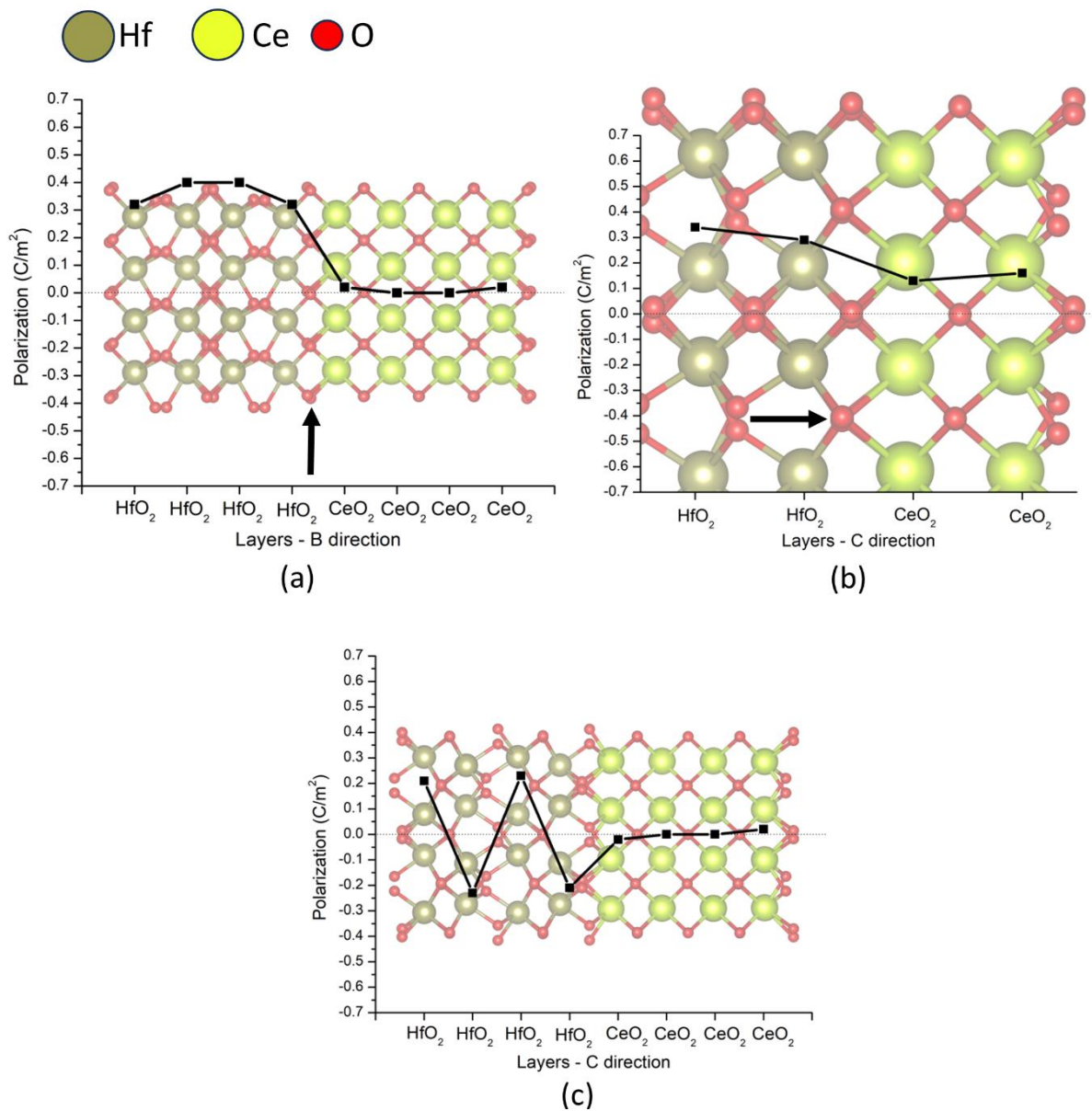

Figure 5. Structure and sublayer polarization of mixed Ce/Hf SLs with (a) 4/4 thickness and polarization in plane (B-SL); (b) 2/2 thickness and (c) 4/4 thickness with polarization out of plane (C-SL). All structures were obtained by relaxing a fully oIII initial configuration.

Finally, the Ce/Hf SLs also relax to mixed configurations, though of a different type. The peculiarities of this composition are preempted by the fact that Ce is the only lanthanide in a list of cations which are otherwise all group IV elements. The $CeO_2$ layer relaxes to a pseudocubic structure (similar to its bulk ground state) in most of the SLs. In the favorable 4/4 B-SLs, the $HfO_2$ layer remains in the polar oIII phase, with a sublayer polarization profile similar to the previously discussed mixed B-SLs (Fig 5a). In the competitive 2/2 C-SL, the $HfO_2$ layer retains the polar structure while inducing a small polarization in the pseudocubic $CeO_2$ layer (Fig. 5b). However, in the 4/4 C-SL, the depolarizing field imposed by the $CeO_2$ layer becomes too strong, and the initially oIII $HfO_2$ layer relaxes into the m phase (Fig 5c). This is accompanied by a significant reduction in formation energy, and this mixed cubic/monoclinic SL proves to be the lowest energy configuration of the 4/4 SLs. However, we find that this relaxation can be avoided – and an out-of-plane polarization of the $HfO_2$ layer retained – by reducing the thickness of the $CeO_2$ layer relative to the $HfO_2$ layer, as seen in the results for the 2/4 c/oIII Ce/Hf C-SL shown in Supplementary Fig S2.

The least energetically viable composition for these mixed SLs is Si/Hf, where both the initially oIII and m SLs undergo large structural distortions on relaxation, and in several cases lead to high energy configurations that seem unlikely to occur in reality. Of all the compared oxides, the difference in cation radii (and lattice mismatch) is the largest for $SiO_2$ and $HfO_2$, leading to rather unstable systems. Nevertheless, the 4/4 B-SLs provide a competitive SL with a mixed o-AP/oIII configuration similar to Ge/Hf and Ti/Hf, with a similar sublayer polarization profile.

## 3.3. Other promising polar and anti-polar superlattices

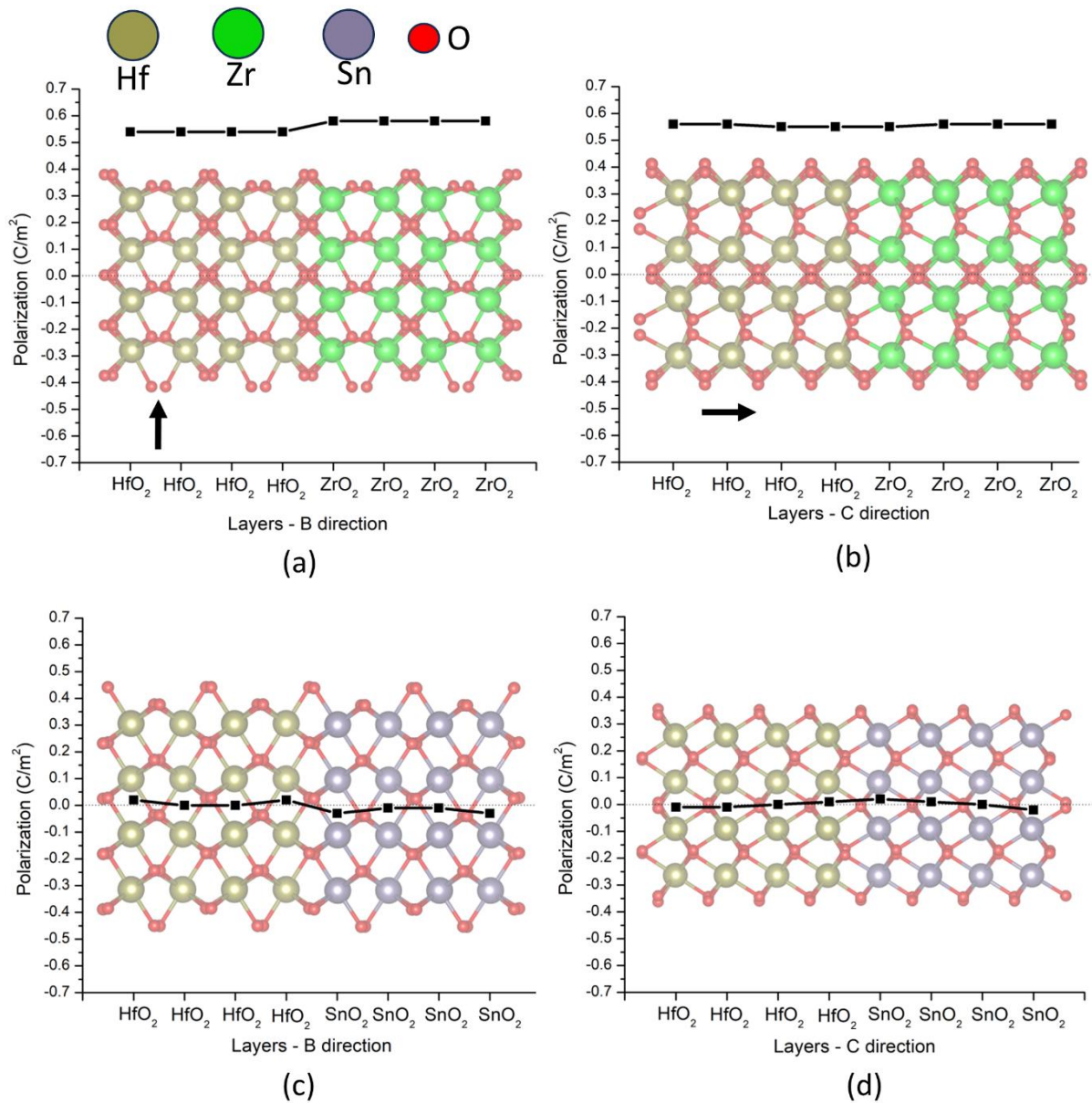

Figure 6. Structure and sublayer polarization of fully polar 4/4 Zr/Hf oIII superlattices with (a) polarization in plane (B-SL) and (b) polarization out of plane (C-SL); and of fully antipolar 4/4 Sn/Hf o-AP superlattices stacked along (c) B-, (d) C-directions. All structures were obtained by relaxing a fully oIII initial configuration.

Energetically competitive structures are obtained for the fully polar Zr/Hf and Pb/Hf SLs, where the m ground state is less dominant than in bulk $HfO_2$. Here, the polarization is always close to that of bulk $HfO_2$ and largely uniform throughout the SL, with only a small kink at the interface (Fig. 6). For all stacking directions and thicknesses, oIII SLs are energetically competitive, with Pb/Hf SLs having a somewhat reduced energy penalty compared to Zr/Hf. This is curious because unlike $ZrO_2$, $PbO_2$ does not have oIII or m local minima, but follows the $HfO_2$ layer, either into the oIII or the m phase, through a strain-induced relaxation (which we also obtain in the bulk compound when imposing SL-like strain constraints).

In sharp contrast to the previous examples, the Sn/Hf SLs are almost all favorable, but fully antipolar. Apart from the relatively higher energy 4/4 A-SL, which has a mixed structure, the other SLs starting from the oIII configurations all relax to a fully antipolar o-AP state. Here, SnO$_2$, which has no bulk oIII local minimum and a low energy o-AP phase, drives the HfO$_2$ layer into the o-AP structure through a strain-inudced relaxation (which, similar to PbO$_2$ described above, we can obtain in the bulk compound by imposing SL-like elastic constraints). Interestingly, this fully antipolar phase is more stable in the SLs than in the individual bulk oxides, and can offer a way to stabilize a potentially antiferroelectric state. However, SnO$_2$ has a rutile ground state, and the fully rutile Sn/Hf SLs have energies in the same range as the fully o-AP SLs (Supplementary table S6), which suggests the possibility of competing phases in these materials.

### 3.4. Discussion

The results in the previous sections encompass a large variety of configurations, with seemingly uncorrelated behavior, arising in large part from the varied chemistry and complex polymorphism of these oxides. However a closer look at the energetics of the various bulk polymophs, as well as the formation energy decomposition of the SLs, reveals a clear underlying mechanism responsible for stabilzing non-monoclinic structures.

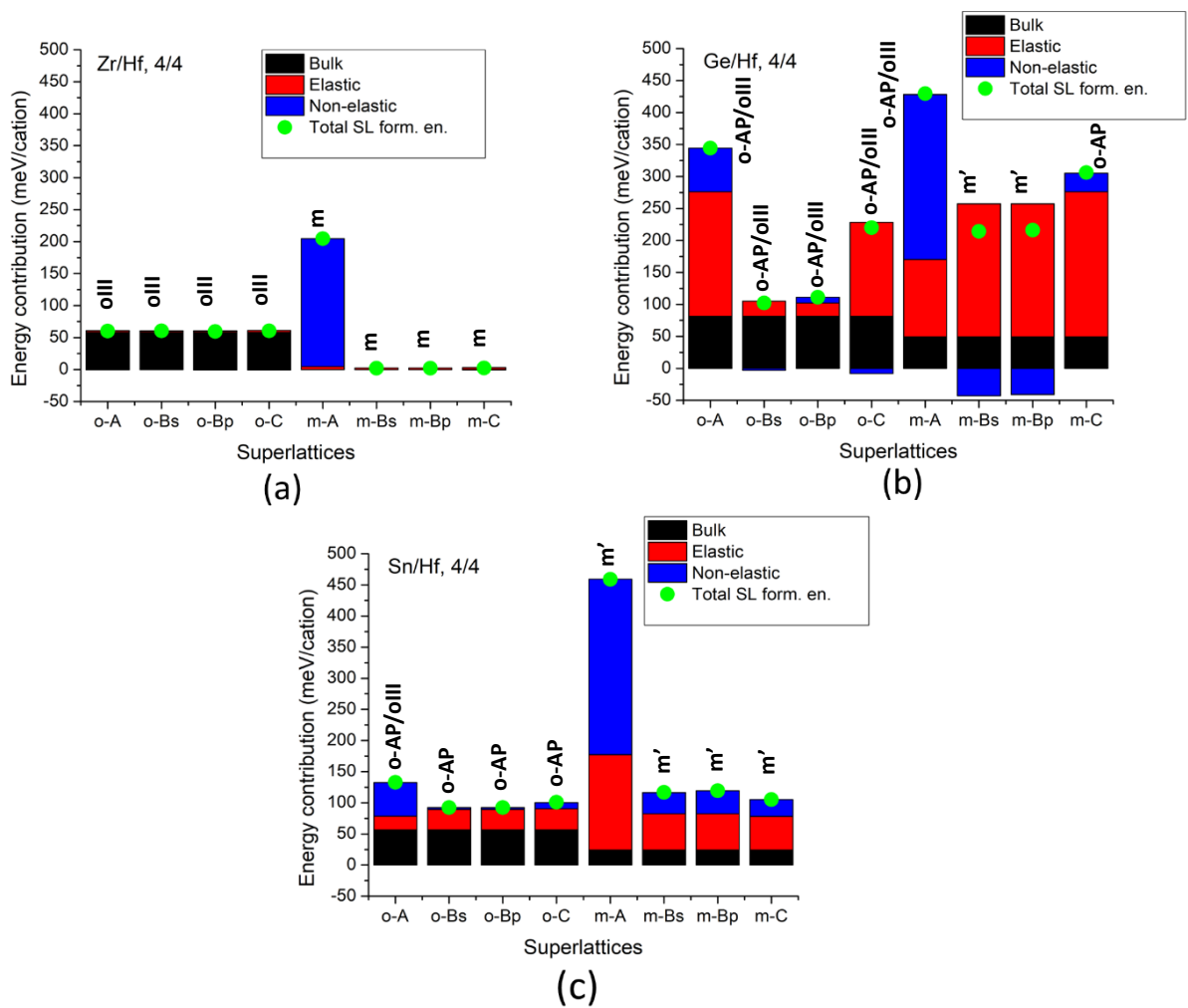

Figure 7. Energy decomposition of Zr/Hf (a), Ge/Hf (b) and Sn/Hf (c) 4/4 SLs. The horizontal axis identifies the initial configurations – o (oIII) and m – as well as the stacking directions – A, C, Bs (B$_{spc}$) and Bp (B$_{pol}$) – while the vertical axis corresponds to the contributions to the formation energy from bulk (black), elastic (red) and non-elastic (blue) effects. The green dots show the total SL formation energy. The labels above each column identify the final structure of the respective SL after relaxation.

We start with the homogenous Zr/Hf SLs, where both bulk oxides have monoclinic ground states and polar local minima of higher energy. Accordingly, the m SLs have lower formation energies than the polar SLs. Since the lattice mismatch between bulk $HfO_2$ and $ZrO_2$ (for both m and oIII phases) is small, the elastic penalties are negligible and the formation energy of the SLs are almost entirely dominated by the bulk contribution (Fig 7a). Accordingly, these fully polar SLs still suffer from large energy penalties – the Zr/Hf oIII SLs have slightly lower energies than bulk oIII $HfO_2$ only because of the relatively low energy of bulk oIII $ZrO_2$.

To instead understand how we obtain favorable SLs, we inspect the representative case of the 4/4 o-AP/oIII Ge/Hf system. In this case, the polar phase with the lowest energy penalty corresponds to the stacking along the B-direction. As can be seen from the formation energy decomposition (Fig 7b), this is because the elastic penalties for the orthorhombic SLs are very small for that particular stacking (in-plane polarization). This allows the low-energy coexistence of the polar oIII (in the $HfO_2$ layer) and antipolar o-AP (in the $GeO_2$ layer) structures in the mixed SLs (Fig 4a). Instead, the (initially) monoclinic Ge/Hf SLs show large elastic contributions to the formation energy, with A-SLs and C-SLs actually relaxing into globally orthorhombic structures – mixed o-AP/oIII for the former and fully o-AP for the latter. This can be attributed to the absence of an m local minimum in bulk $GeO_2$, which ultimately makes the corresponding SLs either unviable or relatively high in energy (Fig 7b). Indeed, we find a similar situation in $TiO_2$ and $CeO_2$, which suggests that a general necessary condition to stabilize the polar structure in the X/Hf SLs is the absence of monoclinic low-energy polymorph in the $XO_2$ layer. We should note here that the stability of these mixed o-AP/oIII SLs (as predicted for $GeO_2$, $TiO_2$ and $CeO_2$) is still quite remarkable, insofar as neither layer is in its bulk ground state. Rather, it is the structural similarity between the o-AP and oIII structures (in contrast to the bulk monoclinic or tetragonal rutile ground states) that yields the o-AP/oIII mixed ground state for these systems.

The absence of a monoclinic polymorph for the $XO_2$ compound may also lead to non-polar (non-monoclinic) solutions. As an example, we may compare the *fully* antipolar Sn/Hf SLs with the mixed Ge/Hf SLs. Both $GeO_2$ and $SnO_2$ have similar polymorphism in the bulk (Table 1), which results in m-SLs having higher formation energies than the orthorhombic ones (Fig 7c). However, constrained to the lattice parameters of the Sn/Hf SL, the $HfO_2$ layer becomes o-AP and, thus, a polar structure cannot be stabilized (Fig 6c,d). Hence, we can conclude that the absence of a competing monoclinic phase is not a sufficient condition to stabilize polar SLs. Indeed, a second necessary condition is needed, namely, that the strain-state of the SLs supports the polar phase in at least the $HfO_2$ layer.

Finally, it is worth noting that in the mixed nonpolar/polar SLs the lowest energy configurations are obtained with the polarization in the plane of the interface. This clearly resonates with well-known electrostatic effects in ferroelectric/dielectric superlattices (e.g., made by perovskite oxides $PbTiO_3$ and $SrTiO_3$) wherein states with in-plane polarization (and no accumulation of bound charges at the interface) are favored over those with an out-of-plane polarization (which inevitably yields interfacial bound charges and depolarizaing fields)[32]. This may also be the origin of the results obtained for the Ce/Hf C-SLs, where going from a 2/2 (Fig 6b) to a 4/4 thickness (Fig 6c) causes the $HfO_2$ layer to go into the m-phase and the polar structure to be lost. It is also interesting to note the case of the mixed 4/4 Ge/Hf C-SL, where the out-of-plane configuration seems somewhat more robust. To further test the stability of this C-oriented polar solution, we considered a SL where the $HfO_2$ layer displays a competing C-oriented polymorph, namely, the oI structure, which can be thought of as composed of anti-parallel domains of the polar oIII phase (Fig 1d). As compared to oIII, this oI structure presents no net interfacial bound charges due to the interfacial discontinuity with the $GeO_2$ layer, which essentially cancels the depolarizing fields. However, the $GeO_2$ layer does not adapt to the oI structure of the $HfO_2$ layer and becomes distorted (oI', Supplementary Note S1). Accordingly, these SLs suffer larger energy penalties compared to the o-AP/oIII C-SL, and the mixed nonpolar/polar structure prevails.

The sensitivity of the SLs to elastic and electrostatic factors allows us to speculate on a further potential design parameter – the relative thickness of the individual layers. For example, while a 1:1 ratio clearly favors a fully antipolar phase in the Sn/Hf SLs, it can be reasonably expected that continuously increasing the thickness of the $HfO_2$ layer relative to $SnO_2$ will eventually lead to a polar structure. Hence, if we consider fully antipolar SLs just below this critical thickness ratio, we might be able to drive such systems into a polar phase by application of an electric field, which could provide us with an interesting family of materials to optimize antiferroelectric behavior. Similarly, increasing the relative thickness of the $HfO_2$ layer in the Ce/Hf C-SL will tend to favor structures with an out-of-plane polarization (see Supplementary Figure S2), and whose stability – relative to non-polar states – can potentially be optimized. This may allow us to tune the energy barriers for ferroelectric switching and thus, potentially, control (reduce) the coercive fields while preserving a robust remnant polarization.

To conclude this discussion, let us note that, despite the large number of systems studied, the present work should be considered neither exhaustive nor fully conclusive. Indeed, here we have only considered perfect, fully relaxed, monodomain, infinite crystal SLs stacked along specific crystallographic axes – by contrast, the effects of defects, epitaxial strain, domain formation, surfaces, and alternative stacking directions have not been studied. Secondly, our work only addresses the thermodynamic stability of the discussed superlattices, without considering kinetic effects. Crucially however, the novel mixed SLs discussed in this work are generally found to have energy costs lower than Zr/Hf SLs, which have already been synthesized and display promising properties. Hence, many of the SLs considered here – which show a better stability of the ferroelectric phase – are good candidates to improve over the Zr/Hf systems and further optimize performance.

**Conclusions**

In the present study, DFT calculations are used to study the structure and energetics of $HfO_2$-based simple oxide superlattices. Most remarkably, we identify several combinations presenting dominant ferroelectric phases. The necessary conditions favoring the stabilization of polar phases in these superlattices are twofold: (i) the absence of the monoclinic ground state in the non-$HfO_2$ oxide (which drives up the energy of corresponding monoclinic superlattices) and (ii) compatible elastic matching between the layers. The predicted ferroelectric solutions tend to present an in-plane polarization so as to minimize depolarizing fields. In addition, we also find other interesting ground states – e.g., of antipolar nature – that could provide a platform for the optimization of $HfO_2$-based antiferroelectrics.

**Methodology**

The first-principles calculations were performed with density functional theory using the plane-wave basis set as implemented in the Vienna ab-initio simulation package (VASP)[33–35]. The electron exchange correlation functional was approximated using the Perdew-Burke-Ernzerhof (PBE) form of the generalized gradient approximation (GGA) with PBEsol modification[36]. A cutoff of 600 eV was used for the expansion of the plane-waves. The valence states explicitly considered for the different elements are as follows: O – $2s^2$, $2p^2$; Ce – $4f^1$, $5d^1$, $6s^2$; Ge – $3d^{10}$, $4s^2$, $4p^2$; Pb – $5d^{10}$, $6s^2$, $6p^2$; Hf – $5p^6$, $5d^2$, $6s^2$; Si – $3s^2$, $3p^2$; Sn – $4d^{10}$, $5s^2$, $5p^2$; Ti – $3p^6$, $3d^2$, $4s^2$ ; Zr – $4s^2$, $4p^6$, $4d^2$, $5s^2$. For bulk structures, a 4x4x4 k-mesh was used to sample the Brillouin zone, with a proportional reduction to [4x4x2] and [4x4x1] along the stacking direction for the SLs. The structures were relaxed until the Hellmann-Feynman forces on each atom fell below 0.01 eV/Å.

VASPKIT[37] and FINDSYM[38] were used for postprocessing, while VESTA[39] was used for structure visualization.

The layer-by-layer polarization for the SLs was computed as the product of the nominal charges (+4 for cations, -2 for oxygen), and the displacements of the ions with respect to the high symmetry cubic fluorite parent structure (Fm-3m), normalized by the volume of half a single unit cell (i.e. a 'sublayer').

## Acknowledgements

BM would like to thank Dr. Hugo Aramberri for many useful discussions. This work was supported by the Luxembourg National Research Fund though grant Nos. INTER/NOW/20/15079143/TRICOLOR.

## References

1. Schroeder, U., Park, M. H., Mikolajick, T. & Hwang, C. S. The fundamentals and applications of ferroelectric HfO2. *Nature Reviews Materials* vol. 7 653–669 Preprint at https://doi.org/10.1038/s41578-022-00431-2 (2022).

2. Böscke, T. S., Müller, J., Bräuhaus, D., Schröder, U. & Böttger, U. Ferroelectricity in hafnium oxide thin films. *Appl Phys Lett* **99**, (2011).

3. Delodovici, F., Barone, P. & Picozzi, S. Trilinear-coupling-driven ferroelectricity in HfO2. *Phys Rev Mater* **5**, (2021).

4. Zhou, S., Zhang, J. & Rappe, A. M. *Strain-induced antipolar phase in hafnia stabilizes robust thin-film ferroelectricity*. *Sci. Adv* vol. 8 https://www.science.org (2022).

5. Raeliarijaona, A. & Cohen, R. E. Hafnia HfO2 is a proper ferroelectric. *Phys Rev B* **108**, (2023).

6. Aramberri, H. & Íñiguez, J. Theoretical approach to ferroelectricity in hafnia and related materials. *Commun Mater* **4**, (2023).

7. Schroeder, U. *et al.* Temperature-Dependent Phase Transitions in HfxZr1-xO2 Mixed Oxides: Indications of a Proper Ferroelectric Material. *Adv Electron Mater* **8**, (2022).

8. Dutta, S. *et al.* Piezoelectricity in hafnia. *Nat Commun* **12**, (2021).

9. Liu, J., Liu, S., Liu, L. H., Hanrahan, B. & Pantelides, S. T. Origin of Pyroelectricity in Ferroelectric HfO2. *Phys Rev Appl* **12**, (2019).

10. Liu, J., Liu, S., Yang, J. Y. & Liu, L. Electric Auxetic Effect in Piezoelectrics. *Phys Rev Lett* **125**, (2020).

11. Wu, Y. *et al.* Unconventional Polarization-Switching Mechanism in $ZrO_2$ Ferroelectrics and Its Implications. *Phys Rev Lett* **131**, 226802 (2023).

12. Qi, Y., Reyes-Lillo, S. E. & Rabe, K. M. 'Double-path' ferroelectrics and the sign of the piezoelectric response. (2022).

13. Materlik, R., Kunneth, C. & Kersch, A. The origin of ferroelectricity in Hf1-xZrxO2: A computational investigation and a surface energy model. *J Appl Phys* **117**, (2015).

14. Yuan, J. H. *et al.* Ferroelectricity in HfO2 from a Coordination Number Perspective. *Chemistry of Materials* **35**, 94–103 (2023).

15. *Ferroelectricity in Doped Hafnium Oxide: Materials, Properties and Devices*. (Elsevier, 2019). doi:10.1016/C2017-0-01145-X.

16. Park, M. H., Lee, Y. H. & Hwang, C. S. Understanding ferroelectric phase formation in doped HfO2 thin films based on classical nucleation theory. *Nanoscale* **11**, 19477–19487 (2019).

17. Xu, X. *et al.* Kinetically stabilized ferroelectricity in bulk single-crystalline HfO2:Y. *Nat Mater* **20**, 826–832 (2021).

18. Materano, M. *et al.* Interplay between oxygen defects and dopants: Effect on structure and performance of HfO2-based ferroelectrics. *Inorganic Chemistry Frontiers* vol. 8 2650–2672 Preprint at https://doi.org/10.1039/d1qi00167a (2021).

19. Dutta, S., Aramberri, H., Schenk, T. & Íñiguez, J. Effect of Dopant Ordering on the Stability of Ferroelectric Hafnia. *Physica Status Solidi - Rapid Research Letters* **14**, (2020).

20. Lehninger, D. *et al.* Ferroelectric [HfO 2 /ZrO 2 ] Superlattices with Enhanced Polarization, Tailored Coercive Field, and Improved High Temperature Reliability . *Advanced Physics Research* **2**, (2023).

21. Liang, Y. K. *et al.* ZrO2-HfO2Superlattice Ferroelectric Capacitors with Optimized Annealing to Achieve Extremely High Polarization Stability. *IEEE Electron Device Letters* **43**, 1451–1454 (2022).

22. Cheema, S. S. *et al.* Ultrathin ferroic HfO2–ZrO2 superlattice gate stack for advanced transistors. *Nature* **604**, 65–71 (2022).

23. Bai, N. *et al.* Designing Wake-Up Free Ferroelectric Capacitors Based on the HfO2/ZrO2 Superlattice Structure. *Adv Electron Mater* **9**, (2023).

24. Park, M. H. *et al.* A comprehensive study on the mechanism of ferroelectric phase formation in hafnia-zirconia nanolaminates and superlattices. *Appl Phys Rev* **6**, (2019).

25. Gong, Z. *et al.* Physical origin of the endurance improvement for HfO2-ZrO2 superlattice ferroelectric film. *Appl Phys Lett* **121**, (2022).

26. Shannon, R. D. *Revised Effective Ionic Radii and Systematic Studies of Interatomie Distances in Halides and Chaleogenides*. Acta Cryst vol. 32 (1976).

27. Aamlid, S. S. *et al.* Phase stability of entropy stabilized oxides with the α-PbO2 structure. *Commun Mater* **4**, (2023).

28. Zhu, T. & Gao, S. P. The stability, electronic structure, and optical property of tio 2 polymorphs. *Journal of Physical Chemistry C* **118**, 11385–11396 (2014).

29. Deringer, V. L., Lumeij, M., Stoffel, R. P. & Dronskowski, R. Ab initio study of the high-temperature phase transition in crystalline GeO2. *J Comput Chem* **34**, 2320–2326 (2013).

30. Mazumder, J. T., Mayengbam, R. & Tripathy, S. K. Theoretical investigation on structural, electronic, optical and elastic properties of TiO2, SnO2, ZrO2 and HfO2 using SCAN meta-GGA functional: A DFT study. *Mater Chem Phys* **254**, (2020).

31. Coh, S. & Vanderbilt, D. Structural stability and lattice dynamics of SiO2 cristobalite. *Phys Rev B Condens Matter Mater Phys* **78**, (2008).

32. Junquera, J. *et al.* Topological phases in polar oxide nanostructures. *Rev Mod Phys* **95**, (2023).

33. Kresse, G. & Furthmüller, J. Efficiency of ab-initio total energy calculations for metals and semiconductors using a plane-wave basis set. *Comput Mater Sci* **6**, 15–50 (1996).

34. Kresse, G. & Furthmüller, J. Efficient iterative schemes for ab initio total-energy calculations using a plane-wave basis set. *Phys Rev B Condens Matter Mater Phys* **54**, 11169–11186 (1996).

35. Kresse, G. & Hafner, J. Ab initio molecular-dynamics simulation of the liquid-metalamorphous- semiconductor transition in germanium. *Phys Rev B* **49**, 14251–14269 (1994).

36. Terentjev, A. V., Constantin, L. A. & Pitarke, J. M. Dispersion-corrected PBEsol exchange-correlation functional. *Phys Rev B* **98**, 1–12 (2018).

37. Wang, V., Xu, N., Liu, J. C., Tang, G. & Geng, W. T. VASPKIT: A user-friendly interface facilitating high-throughput computing and analysis using VASP code. *Comput Phys Commun* **267**, 108033 (2021).

38. Stokes, H. T. & Hatch, D. M. FINDSYM: Program for identifying the space-group symmetry of a crystal. *Journal of Applied Crystallography* vol. 38 237–238 Preprint at https://doi.org/10.1107/S0021889804031528 (2005).

39. Momma, K. & Izumi, F. VESTA 3 for three-dimensional visualization of crystal, volumetric and morphology data. *J Appl Crystallogr* **44**, 1272–1276 (2011).

# Supplementary Information for "First-principles predictions of HfO$_2$-based ferroelectric superlattices"

Binayak Mukherjee[1], Natalya S. Fedorova[1] and Jorge Íñiguez-González[1,2]

[1]Department of Materials Research and Technology, Luxembourg Institute of Science and Technology, L-4362, Esch-sur-Alzette, Luxembourg

[2]Department of Physics and Materials Science, University of Luxembourg, L-4422, Belvaux, Luxembourg

**Table S1**. Initial structure, relaxed structure, and formation energies for Ce/Hf SLs. Phase labels as explained in the main text.

|  | 2/2 | | 4/4 | |
|---|---|---|---|---|
| Initial structure (stacking) | Form. En. (meV/cation) | Relaxed structure | Form. En. (meV/cation) | Relaxed structure |
| oIII (A) | 230 | c/oIII | 222 | c/oIII |
| oIII (B$_{spc}$) | 158 | c/o-AP | 148 | c/oIII |
| oIII (B$_{pol}$) | 151 | c/o-AP | 132 | c/oIII |
| oIII (C) | 184 | c/oIII | 103 | c/m |
| m (A) | 158 | c/o-AP | 281 | c/o-AP |
| m (B$_{spc}$) | 277 | m | 261 | c/m |
| m (B$_{pol}$) | 153 | m | 209 | m' |
| m (C) | 294 | m | 178 | c/m' |
| t (A) | 202 | t | 179 | t |
| t (B) | 202 | t | 179 | t |
| t (C) | 260 | t | 244 | t |
| oI (A) | 305 | oI | 312 | oI |
| oI (B) | - | - | 294 | oI |
| oI (C) | 200 | oI | 173 | oI |

**Table S2.** Initial structure, relaxed structure, and formation energies for Ge/Hf SLs. Phase labels as explained in the main text.

|  | 2/2 | | 4/4 | |
|---|---|---|---|---|
| Initial structure (stacking) | Form. En. (meV/cation) | Relaxed structure | Form. En. (meV/cation) | Relaxed structure |
| oIII (A) | 201 | o-AP | 344 | o-AP/oIII |
| oIII (B$_{spc}$) | 84 | o-AP/oIII | 102 | o-AP/oIII |
| oIII (B$_{pol}$) | 84 | o-AP/oIII | 111 | o-AP/oIII |
| oIII (C) | 155 | o-AP/oIII | 220 | o-AP/oIII |
| m (A) | 84 | o-AP/oIII | 429 | o-AP/oIII |
| m (B$_{spc}$) | 148 | m' | 214 | m' |
| m (B$_{pol}$) | 148 | m' | 216 | m' |
| m (C) | 319 | o-AP | 306 | o-AP |
| t (A) | 333 | t | 431 | t |
| t (B) | 333 | t | 431 | t |
| t (C) | 744 | t | 710 | t |
| oI (A) | 590 | oI' | 526 | oI' |
| oI (B) | - | - | 413 | oI' |
| oI (C) | 190 | oI' | 361 | oI' |
| Rutile (B) | 240 | Rutile | 242 | Rutile |
| Rutile (C) | 257 | Rutile | 240 | Rutile |
| Rutile/m (A) | - | - | 235 | Rutile/m |
| Rutile/m (B) | - | - | 297 | Rutile/m |
| Rutile/m (C) | - | - | 1543 | Rutile/m |

**Table S3.** Initial structure, relaxed structure, and formation energies for Pb/Hf SLs. Phase labels as explained in the main text.

| | 2/2 | | 4/4 | |
|---|---|---|---|---|
| Initial structure (stacking) | Form. En. (meV/cation) | Relaxed structure | Form. En. (meV/cation) | Relaxed structure |
| oIII (A) | 127 | oIII | 120 | oIII |
| oIII ($B_{spc}$) | 140 | oIII | 128 | oIII |
| oIII ($B_{pol}$) | 108 | oIII | 113 | oIII |
| oIII (C) | 121 | oIII | 119 | oIII |
| m (A) | 129 | m | 337 | m |
| m ($B_{spc}$) | 103 | m | 92 | m |
| m ($B_{pol}$) | 109 | m | 95 | m |
| m (C) | 95 | m' | 74 | m' |
| t (A) | 208 | t | 205 | t |
| t (B) | 207 | t | 205 | t |
| t (C) | 163 | t | 174 | t |
| oI (A) | 120 | oI | 117 | oI |
| oI (B) | - | - | 111 | oI |
| oI (C) | 112 | oI | 107 | oI |

**Table S4.** Initial structure, relaxed structure, and formation energies for Si/Hf SLs. Phase labels as explained in the main text.

| | 2/2 | | 4/4 | |
|---|---|---|---|---|
| Initial structure (stacking) | Form. En. (meV/cation) | Relaxed structure | Form. En. (meV/cation) | Relaxed structure |
| oIII (A) | 489 | o-AP | 772 | o-AP/oIII |
| oIII ($B_{spc}$) | 242 | o-AP | 324 | o-AP/oIII |
| oIII ($B_{pol}$) | 242 | o-AP | 307 | o-AP/oIII |
| oIII (C) | 406 | o-AP/oIII | 544 | o-AP/oIII |
| m (A) | 242 | o-AP/c | 658 | m' |
| m (B$_{spc}$) | 411 | m' | 545 | m' |
| m (B$_{pol}$) | 411 | m' | 547 | m' |
| m (C) | 355 | m' | 266 | m' |
| t (A) | 543 | t | 754 | t |
| t (B) | 543 | t | 754 | t |
| t (C) | 1433 | t | 1406 | t |
| oI (A) | 432 | oI' | 303 | oI' |
| oI (B) | - | oI' | 333 | oI' |
| oI (C) | 379 | oI' | 324 | oI' |

**Table S5.** Initial structure, relaxed structure, and formation energies for Sn/Hf SLs. Phase labels as explained in the main text.

| | 2/2 | | 4/4 | |
|---|---|---|---|---|
| Initial structure (stacking) | Form. En. (meV/cation) | Relaxed structure | Form. En. (meV/cation) | Relaxed structure |
| oIII (A) | 96 | o-AP | 133 | o-AP/oIII |
| oIII (B$_{spc}$) | 95 | o-AP | 92 | o-AP |
| oIII (B$_{pol}$) | 95 | o-AP | 92 | o-AP |
| oIII (C) | 104 | o-AP | 101 | o-AP |
| m (A) | 95 | o-AP | 459 | m' |
| m (B$_{spc}$) | 129 | m' | 117 | m' |
| m (B$_{pol}$) | 142 | m' | 120 | m' |
| m (C) | 130 | m' | 105 | m' |
| t (A) | 299 | t | 299 | t |
| t (B) | 299 | t | 299 | t |
| t (C) | 292 | t | 292 | t |
| oI (A) | 213 | oI' | 202 | oI' |
| oI (B) | - | - | 203 | oI' |
| oI (C) | 208 | oI' | 198 | oI' |
| Rutile (B) | 90 | Rutile | 90 | Rutile |
| Rutile (C) | 128 | Rutile | 109 | Rutile |

**Table S6.** Initial structure, relaxed structure, and formation energies for Ti/Hf SLs. Phase labels as explained in the main text.

| | 2/2 | | 4/4 | |
|---|---|---|---|---|
| Initial structure (stacking) | Form. En. (meV/cation) | Relaxed structure | Form. En. (meV/cation) | Relaxed structure |
| oIII (A) | 156 | oIII | 186 | oIII |
| oIII (B$_{spc}$) | 46 | o-AP/oIII | 59 | o-AP/oIII |
| oIII (B$_{pol}$) | 46 | o-AP/oIII | 73 | o-AP/oIII |
| oIII (C) | 79 | o-AP/oIII | 88 | o-AP/oIII |
| m (A) | 67 | m | 282 | m |
| m ($B_{spc}$) | 59 | m | 56 | m |
| m ($B_{pol}$) | 69 | m | 67 | m |
| m (C) | 85 | m | 83 | m |
| t (A) | 194 | t | 221 | t |
| t (B) | 194 | t | 221 | t |
| t (C) | 340 | t | 326 | t |
| oI (A) | 162 | oI | 165 | oI |
| oI (B) | - | - | 161 | oI |
| oI (C) | 168 | oI | 173 | oI |

**Table S7.** Initial structure, relaxed structure, and formation energies for Zr/Hf SLs. Phase labels as explained in the main text.

| | 2/2 | | 4/4 | |
|---|---|---|---|---|
| Initial structure (stacking) | Form. En. (meV/cation) | Relaxed structure | Form. En. (meV/cation) | Relaxed structure |
| oIII (A) | 60 | oIII | 60 | oIII |
| oIII ($B_{spc}$) | 61 | oIII | 60 | oIII |
| oIII ($B_{pol}$) | 60 | oIII | 60 | oIII |
| oIII (C) | 56 | oIII | 60 | oIII |
| m (A) | 2 | m | 205 | m |
| m ($B_{spc}$) | 2 | m | 2 | m |
| m ($B_{pol}$) | 2 | m | 2 | m |
| m (C) | 2 | m | 2 | m |
| t (A) | 111 | t | 111 | t |
| t (B) | 111 | t | 111 | t |
| t (C) | 112 | t | 112 | t |
| oI (A) | 45 | oI | 45 | oI |
| oI (B) | - | oI | 46 | oI |
| oI (C) | 46 | oI | 46 | oI |

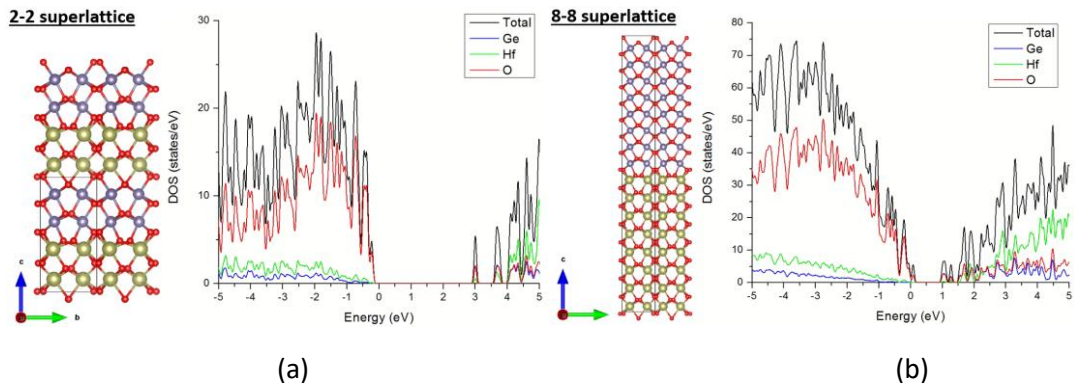

**Figure S1.** Electronic density of states for a 2/2 and 8/8 mixed o-AP/oIII Ge/Hf C-SL, with polarization out-of-plane. A reduction in the band gap is observed with increasing layer thickness, due to increased charged confinement at the polar/nonpolar interface. A metallic state however is avoided up to 8-8 thickness. It should be noted that the gap is calculated with a PBEsol functional, which is known to underestimate the band gap.

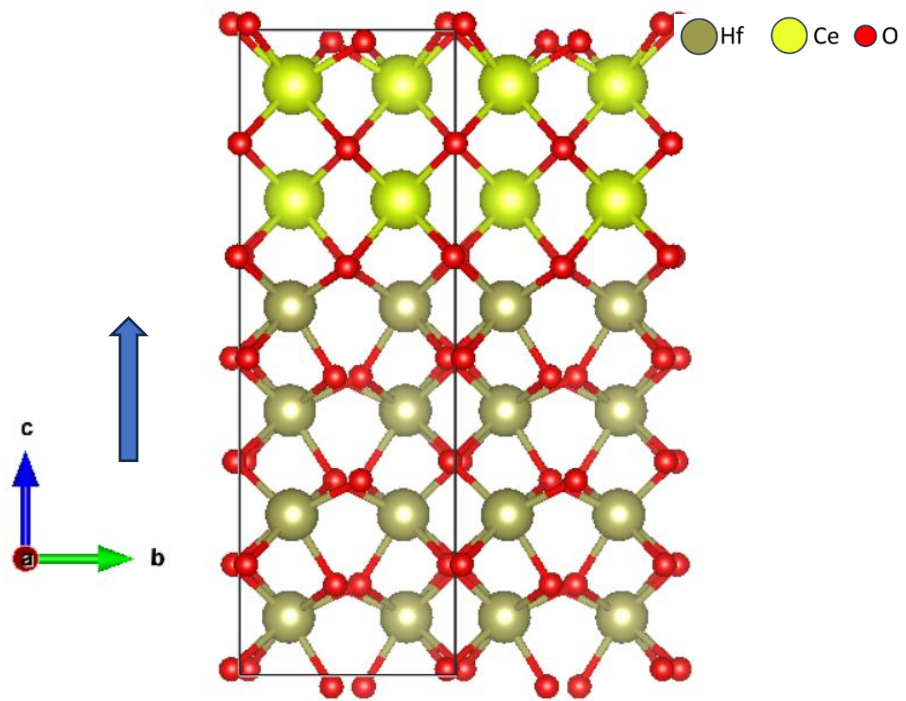

**Figure S2.** 2/4 Ce/Hf C-SL, with polarization parallel to the stacking direction. The arrow indicates the direction of polarization.

**Supplementary Note S1:** Antipolar oI vs mixed antipolar/polar o-AP/oIII Ge/Hf C-SLs

When starting from an oI geometry, bulk $GeO_2$ relaxes into a structure where the cation coordination number tends towards 6, in contrast to $HfO_2$ where the oI structure has a cation coordination number tending towards 7. We refer to the distorted bulk structure of $GeO_2$ as oI'.

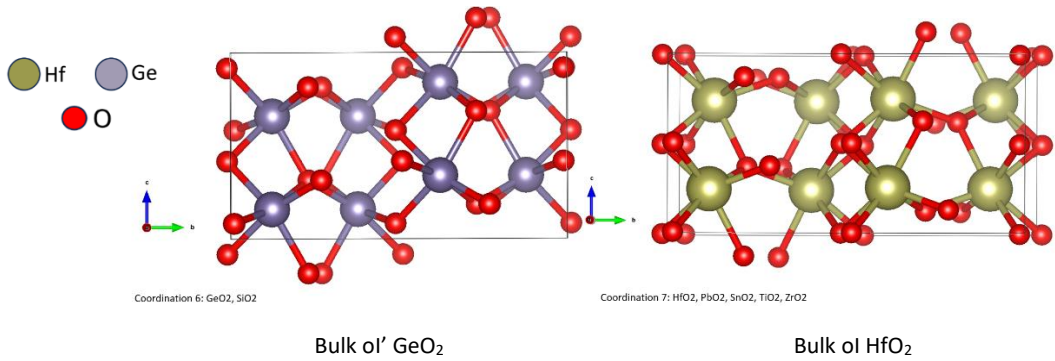

Bulk oI' GeO$_2$            Bulk oI HfO$_2$

While both structures have the Pbca space group, the bulk oI' phase has a significantly higher energy compared to oI (see Table 1, main text). Accordingly, the GeO$_2$ layer in the antipolar Ge/Hf SLs relaxes to oI', and the resulting SL has a much higher formation energy than the competitive antipolar/polar o-AP/oIII C-SL (Fig 4b, main text).

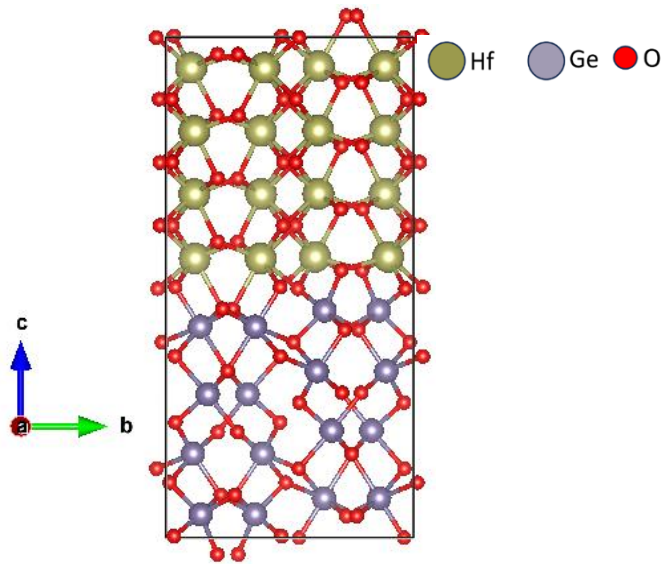

High energy fully antipolar oI/oI' C-SL

Since the starting phase in the SL above is very high energy for bulk GeO$_2$, an argument can be made that this starting condition disfavors the antipolar solution. To avoid this possibility, we consider another alternative antipolar starting condition where GeO$_2$ is in its favored o-AP phase while HfO$_2$ is oI.

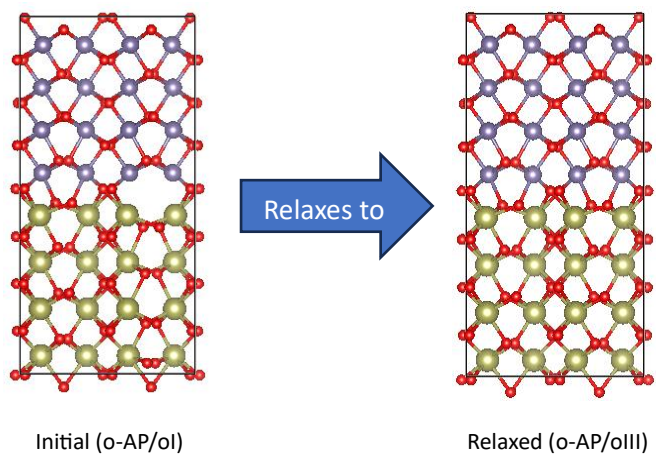

Initial (o-AP/oI)          Relaxes to          Relaxed (o-AP/oIII)

As seen above, this mixed starting configuration is not a local energy minimum, and the HfO$_2$ layer relaxes to the polar oIII phase, leading to the mixed o-AP/oIII C-SL. This guarantess that the Ge/Hf SL with out-of-plane polarization is lower in energy than possible fully antipolar configurations.